\documentclass[conference]{IEEEtran}
\IEEEoverridecommandlockouts
\usepackage{cite}
\usepackage{amsmath,amssymb,amsfonts}
\usepackage{algorithmic}
\usepackage{graphicx}
\usepackage{textcomp}
\usepackage{xcolor}
\usepackage{multirow}
\def\BibTeX{{\rm B\kern-.05em{\sc i\kern-.025em b}\kern-.08em
    T\kern-.1667em\lower.7ex\hbox{E}\kern-.125emX}} 
\begin{document}
\title{Choose your tools carefully: A Comparative Evaluation of Deterministic vs. Stochastic and Binary vs. Analog Neuron models for Implementing Emerging Computing Paradigms
\thanks{\IEEEauthorrefmark{1}mm8by@virginia.edu\vspace{-1ex},\IEEEauthorrefmark{2}gangulys2@vcu.edu\vspace{-1ex}}}
\author{\fontsize{11}{11}\selectfont Md Golam Morshed\textsuperscript{1}\IEEEauthorrefmark{1}, Samiran Ganguly\textsuperscript{2}\IEEEauthorrefmark{2}, and Avik W. Ghosh\textsuperscript{1,3}\\
\fontsize{10}{12}\selectfont \textsuperscript{1}Department of Electrical and Computer Engineering, University of Virginia, Charlottesville, VA, USA\\ \textsuperscript {2}Department of Electrical and Computer Engineering, Virginia Commonwealth University, Richmond, VA, USA\\\textsuperscript {3}Department of Physics, University of Virginia, Charlottesville, VA, USA
}
\maketitle
\begin{abstract}
Neuromorphic computing, commonly understood as a computing approach built upon neurons, synapses, and their dynamics, as opposed to Boolean gates, is gaining large mindshare due to its direct application in solving current and future computing technological problems, such as smart sensing, smart devices, self-hosted and self-contained devices, artificial intelligence (AI) applications, etc. In a largely software-defined implementation of neuromorphic computing, it is possible to throw enormous computational power or optimize models and networks depending on the specific nature of the computational tasks. However, a hardware-based approach needs the identification of well-suited neuronal and synaptic models to obtain high functional and energy efficiency, which is a prime concern in size, weight, and power (SWaP) constrained environments. In this work, we perform a study on the characteristics of hardware neuron models (namely, inference errors, generalizability and robustness, practical implementability, and memory capacity) that have been proposed and demonstrated using a plethora of emerging nano-materials technology-based physical devices, to quantify the performance of such neurons on certain classes of problems that are of great importance in real-time signal processing like tasks in the context of reservoir computing. We find that the answer on which neuron to use for what applications depends on the particulars of the application requirements and constraints themselves, i.e., we need not only a hammer but all sorts of tools in our tool chest for high efficiency and quality neuromorphic computing.
\end{abstract}


\section{Introduction}
High-performance computing has historically developed around the Boolean computing paradigm, executed on silicon (Si) complementary metal oxide semiconductor (CMOS) hardware. In fact, software has for decades been developed around the CMOS fabric that has singularly dictated our choice of materials, devices, circuits, and architecture – leading to the dominant processor design paradigm: von Neumann architecture that separates memory and processing units. Over the last decade, however, Moore’s law for hardware scaling has significantly slowed down, primarily due to the prohibitive energy cost of computing and an increasingly steep memory wall. At the same time, software development has significantly evolved around “Big Data” paradigm, with machine learning and artificial intelligence (AI) dominating the roost. Additionally, the push towards the internet of things (IoT) edge devices has prompted an intensive search for energy-efficient and compact hardware systems for on-chip data processing~\cite{Nature_article}. 

One such direction is neuromorphic computing, which uses the concept of mimicking a human brain architecture to design circuits and systems that can perform highly energy-efficient computations~\cite{neuromorphic_main, survey_neuromorphic, Physics_neuromorphic,2022_roadmap,incorvia}. A human brain is primarily composed of two functional elemental units – synapses and neurons. Neurons are interconnected through synapses with different connection strengths (commonly known as synaptic weights), which provide the learning and memory capabilities of the brain. A neuron receives synaptic inputs from other neurons, generates output in the form of action potentials, and distributes the output to the subsequent neurons. A human brain has $\sim 10^{11}$ neurons and $\sim 10^{15}$ synapses and consumes $\sim 1-10$~$\mathrm{fJ}$ per synaptic event~\cite{book1, book2, Upadhyay2016May}. 

To emulate the organization and functionality of a human brain, there are many proposals for physical neuromorphic computing systems using memristors~\cite{memristor1, memristor2, memristor3}, spintronics~\cite{spintronics1, spintronics2, spintronics3}, charge-density-wave (CDW) devices~\cite{CDW}, photonics~\cite{photonics1,photonics2}, etc. {In recent years, there has been significant progress in the development of physical neuromorphic hardware, both in academia and industry. The hierarchy of neuromorphic hardware implementation spans from the system level to the device level and all the way down to the level of the material. At the system level, various large-scale neuromorphic computers utilize different approaches – for instance, IBM’s TrueNorth~\cite{Merolla2014Aug}, Intel’s Loihi~\cite{Loihi}, SpiNNaker~\cite{SpiNNaker}, BrainScaleS~\cite{BrainScaleS}, Tianjic chip~\cite{Tianjic}, Neurogrid~\cite{Neurogrid}, etc. They support a broad class of problems ranging from complex to more general computations. At the device level, the most commonly used component is the memristor which can be utilized in synapse and neuron implementations~\cite{Memristor_synapse1,Memristor_synapse2,Memristor_neuron1,Memristor_neuron2}. Memristor crossbars are frequently used to represent synapses in neuromorphic systems~\cite{Memristor_crossbar1,Memristor_crossbar2}. Memristor can also provide stochasticity in the neuron model~\cite{memristor_noise}. Another emerging class of devices for neuromorphic computing is spintronics devices~\cite{spintronics1}. Spintronics devices can be implemented with low energy and high density and are compatible with existing CMOS technology~\cite{Sengupta2016May}. The spintronics devices utilized in neuromorphic computing include spin-torque devices~\cite{Torrejon2017Jul,Roy2014Mar,Sengupta2016Jan}, magnetic domain walls~\cite{neuromorphic_domainWalls1,neuromorphic_domainWalls2,neuromorphic_domainWalls3}, and skyrmions~\cite{neuromorphic_skyrmion1,neuromorphic_skyrmion2}. Optical or photonics devices are also implemented for neurons and synapses in recent years~\cite{photonics1,neuromorphic_photonic1,neuromorphic_photonic2}.} The field is very new and many novel forms of neuron and synaptic devices can be designed to match the mathematical model of neural networks (NNs). Physical neuromorphic computing can implement these functionalities directly in their physical characteristics (I-I, V-V, I-V), which results in highly compact devices that are well-suited for scalable and energy-efficient neuromorphic systems~\cite{Kerem_PRX, Kerem_EDL, Samiran_ASN, Yang2013Jan}. This is critical as current NN-based computing is highly centralized (resident-on and accessed-via cloud) and is energy inefficient because the underlying volatile, often von Neumann, digital Boolean-based system design unit has to emulate inherently analog, mostly non-volatile distributed computing model of neural systems, even if at a simple abstraction level~\cite{Merolla2014Aug}. Recent advances in custom design such as FPGAs~\cite{Wang2018Apr} and more experimental Si FPNAs~\cite{Hasler} have demonstrated that a new form of device design rather than emulation is the way to go, and physical neuromorphic computing based on emerging technology can go a long way to achieve this~\cite{Rajendran2016Mar}.

There is an increased use of noise-as-a-feature rather than a nuisance in NN models~\cite{noise1, noise2, noise3}, and physical neuromorphic computing can provide natural stochasticity, with various noise colors depending on the device physics~\cite{Vincent2015Apr,stochasticity1}. Some prominent areas where stochasticity and noise have been used include training generalizability~\cite{Generalizability}, stochastic sampling~\cite{stochasticsampling}, and recently proposed and coming into prominence, diffusion-based generative models~\cite{Generativemodel}. In all these models, noise plays a fundamental role, i.e., these algorithms do not work without inherent noise.

It is therefore critical to study and analyze the kinds of devices that will be useful to implement physical neuromorphic computing. We understand from neurobiology that there is a large degree of neuron design customization that has developed through evolution to obtain high task-based performance. Similarly, a variety of mathematical models of neurons have been designed in NN literature as well~\cite{survey_neuromorphic,neuron_model1, Samiran_ASN}. It is quite likely that the area of physical neuromorphics will use a variety of device designs rather than the uniformity of NAND gate-based design commonly seen in Boolean-based design, to achieve the true benefits of energy efficiency and scalability brought forth by this paradigm of system design.

In this work, we study a subset of this wide variety of neuron designs that are well-represented and easily available from many proposed physical neuromorphic platforms to understand and analyze their task specialization. In particular, we analyze analog and binary neuron models, including stochasticity in the model, for analog temporal inferencing tasks, and evaluate and compare their performances. We numerically estimate the performance metric normalized means squared error (NMSE), discuss the effect of stochasticity on prediction accuracy vs. robustness, and show the hardware implementability of the models. {Furthermore, we estimate the memory capacity for different neuron models.} Our results suggest that analog stochastic neurons perform better for analog temporal inferencing tasks both in terms of prediction accuracy and hardware implementability. {Additionally, analog neurons show larger memory capacity.} Our findings may provide a potential path forward toward efficient neuromorphic computing.  
\begin{figure*}[htbp]
\begin{center}
\includegraphics[width=.90\textwidth]{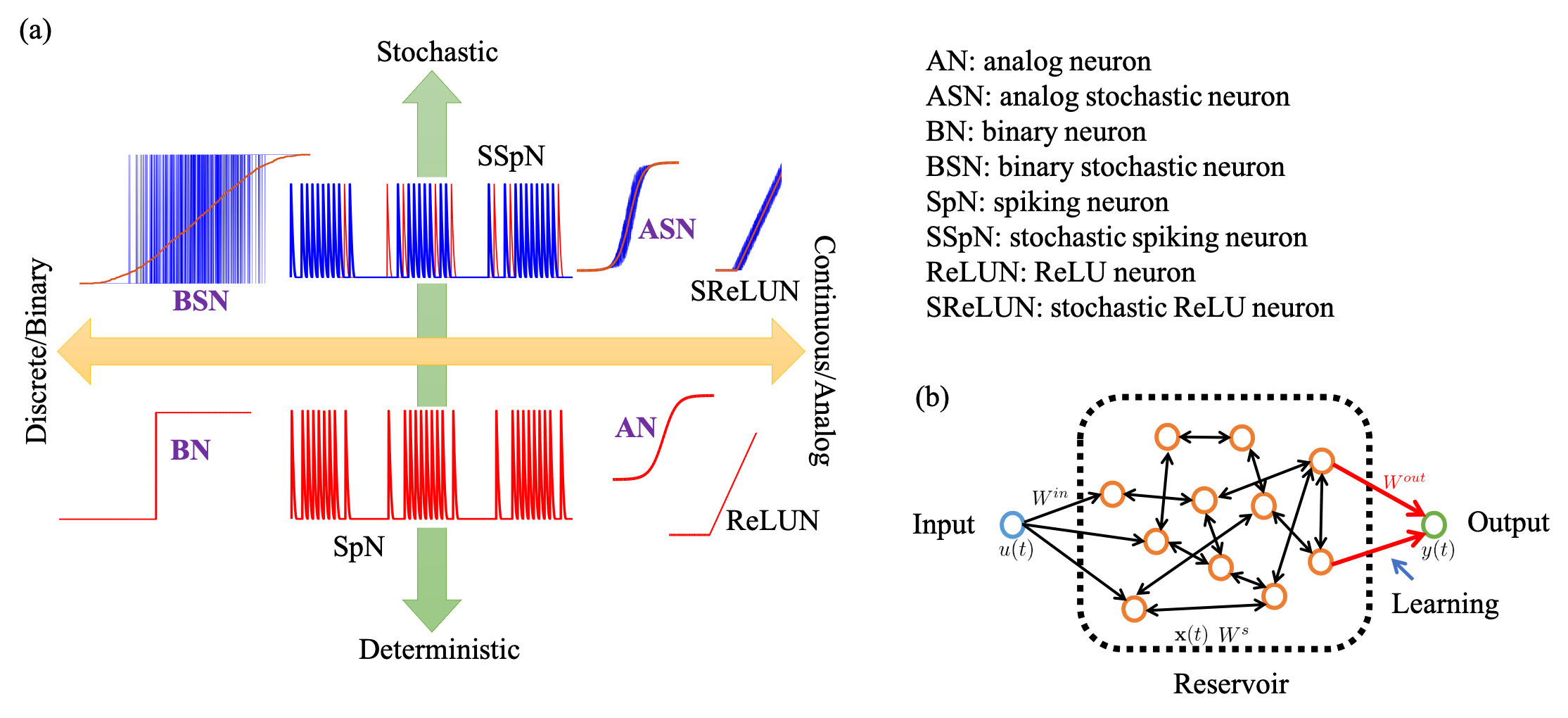}
\end{center}
    \caption{(a) Schematic of different types of widely used neuron models with their output characteristics. In the bottom panel, all the red curves represent the deterministic neurons' output characteristics. In the top panel, the blue curves represent the actual stochastic output characteristics while the red is the corresponding deterministic/expected value of the output ($<stochastic~output>$) characteristics. Spiking neurons (SpN and SSpN) can be considered in between the two limits of purely binary vs. purely analog neurons. Please note that we only analyze the analog and binary neurons (including their stochastic counterparts) in this work, as indicated by the purple-colored bold font labels. (b) Schematic of a reservoir setup using neurons connected with each other bidirectionally with random weights.}
    \label{fig1}
\end{figure*}

\section{Brief Overview on Neuron Models}
An essential function of a neuron in a NN is processing the weighted synaptic inputs and generating an output response. A single biological neuron itself is a complex dynamical system~\cite{biological_neuron}. Proposed artificial neurons in most implementations of NNs (either software or hardware) are significantly simpler unless they specifically attempt to mimic the biological neuron~\cite{artificial_neuron, survey_neuromorphic,Schuman2022Jan}. As such their mathematical representations are cheaper and a significant amount of computational capabilities derive from the network itself. However, a NN is an interplay of the neurons, the synapses, and the network structure itself, and therefore the neuron model itself may provide certain capabilities that can help make a more efficient NN, in the context of the application specialization~\cite{NN_survey}. 

The set of behavior over which such neurons can be classified and analyzed is vast and may include spiking vs. non-spiking behavior with associated data representation, deterministic vs. stochastic output response function, discrete (or binary) vs. continuous (or analog) output response function, the particular mathematical model of the output response function itself (e.g., sigmoid, tanh, ReLU), presence or absence of memory states with a neuron, etc~\cite{deepLearning_book,spiking_nonspiking,Deterministic_stochastic}. In the software NN world, specialization of certain neural models and connectivity are well appreciated, as an example sparse vs. dense vs. convolutional layers, or the use of ReLU neurons in the hidden layers vs. sigmoidal, softmax layers at outputs employed in many computer vision tasks~\cite{activation_function,ReLu_sigmoid,ReluvsSigmoid}. Fig.~\ref{fig1}(a) schematically shows the output characteristics of different types of widely used neuron models.

In this work, we have focused on two particular behaviors of neural models that we believe can capture a significant application space, particularly in the domain of lightweight real-time signal processing tasks, and are readily built from emerging materials technology. We specifically look at binary vs. analog and deterministic vs. stochastic neuron output response functions (purple-colored bold font labels in Fig.~\ref{fig1}(a)). We also use them in a reservoir computing (RC)-like context for signal processing tasks for our analysis. Reservoir computing uses the dynamics of a recurrently connected network of neurons to project an input (spatio-)temporal signal onto a high dimensional phase space, which forms the basis of inference, typically via a shallow 1-layer linear transform or a multi-layer feedforward network~\cite{RC1,RC2,RC3,RC4,memristor3}. A schematic of a reservoir is shown in Fig.~\ref{fig1}(b) where the neurons are connected with each other bidirectionally with random weights. Multiple reservoirs may be connected hierarchically for more complex deep RC architecture. RC may be considered as a machine learning analog of an extended Kalman filter where the state space and the observation models are learned and not designed a priori~\cite{RC1}.

Our choice of evaluating these specific behavior differences on an RC-based NN reflects the prominent use-case that is made out for many emerging nano-materials technology-based neuron and synaptic devices, viz. energy-efficient learning, and inference at the edge. These tasks often end up involving temporal or spatio-temporal data processing to extract relevant and actionable information, some examples being anomaly detection~\cite{anomaly_detection}, feature tracking~\cite{feature_tracking}, optimal control~\cite{optimal_control}, and event prediction~\cite{event_prediction}, all of which are well-suited for an RC-based NN. Therefore this testbench forms a great intersection for our analysis.

It should be noted that we do not include spiking neurons in this particular analysis. Spiking neurons have significantly different data encoding (level vs. rate or inter-spike interval encoding) and learning mechanisms (back-propagation or regression vs. spike-time dependent plasticity) that it is hard to disentangle the neuron model itself from demonstrated tasks, therefore we leave such a contrasting analysis of spiking neuron devices with non-spiking variants for a future study.

The neurons are modeled in the following way:

\begin{equation}
 \textbf{y} = f_N(\sum{w^T\textbf{x}}) + r_N
\end{equation}

Here the symbols have the usual meaning, i.e., $\textbf{y}$ is the output activation of the neuron, $f_N$ is the activation function, which is a sigmoidal or hyperbolic tangent for most non-spiking hardware neurons, and $r_N$ is a random sample drawn from a random uniform distribution to represent stochasticity. It is possible to use a ReLU-like activation function or some other distribution for sampling stochasticity, particularly if the hardware neuron shows colored noise behavior, we do not particularize for such details and keep the analysis confined to the most common hardware neuron variants. Therefore, in our analysis, the $r_N$ term is weighed down by an arbitrary factor to mimic the degree of stochasticity displayed by the neuron, and the $f_N$ is either a continuous $tanh()$ for analog neuron or a $sgn(tanh())$ for a binary neuron ($sgn()$ being the signum function).
\begin{figure*}[htbp]
\begin{center}
\includegraphics[width=.75\textwidth]{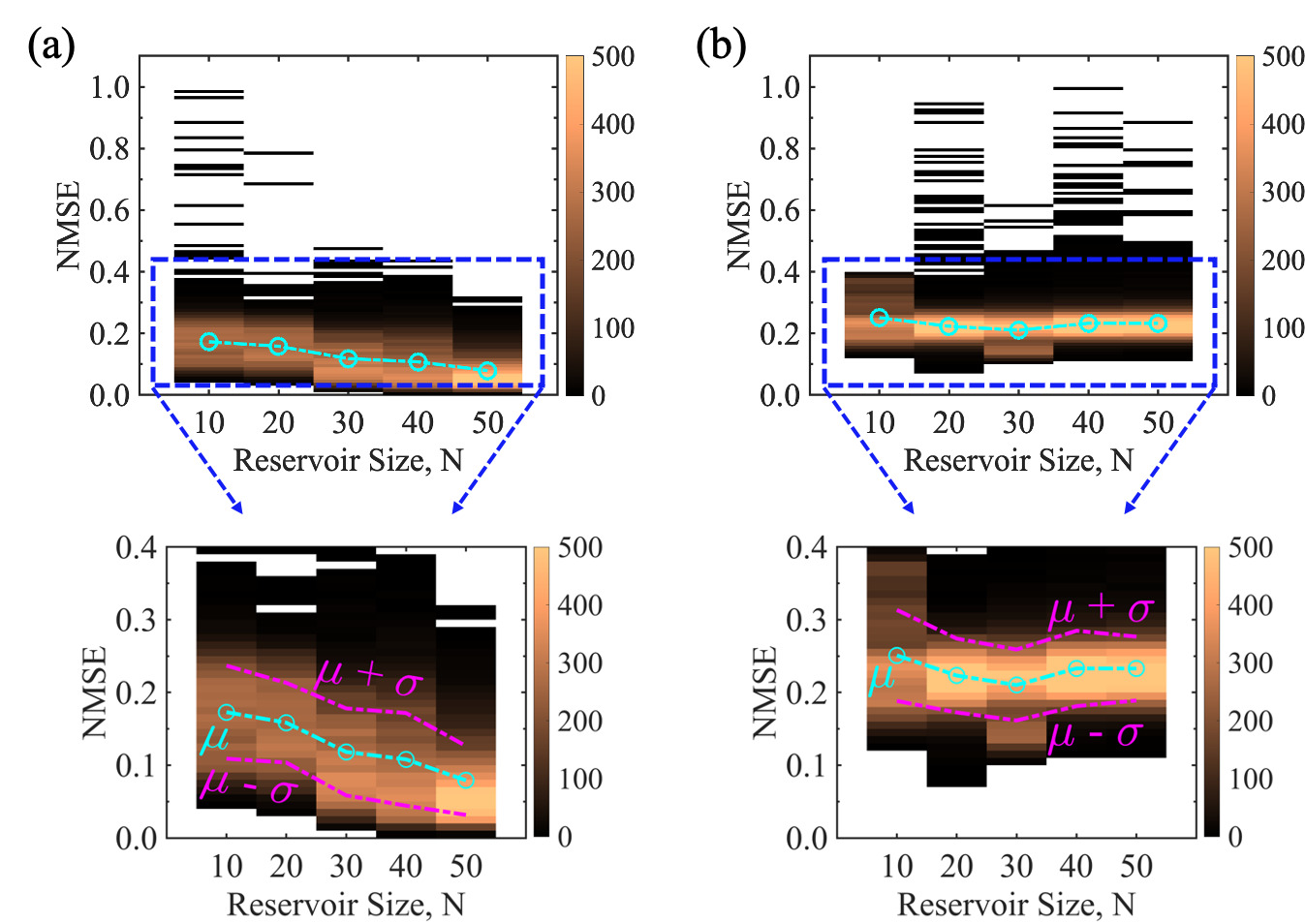}
\end{center}
    \caption{Comparison of NMSE for an analog time-series prediction task between (a) ASN and (b) BSN models as a function of reservoir size with $5\%$ stochasticity incorporated in both the neuron models for a clean input signal. The form of the clean input signal is $u(t)=A \cos(2\pi f_1t)+B \sin(2\pi f_2t)$, where $A=1$, $B=2$, $f_1=0.10~Hz$, and $f_2=0.02~Hz$. ASN performs better than BSN for the entire range of reservoir size as indicated by the average ($\mu$) NMSE (cyan dashed-dotted line). ASN shows a decreasing trend in NMSE as a function of reservoir size while BSN results remain almost unchanged. The NMSE data for every reservoir size is obtained from five different reservoir topologies and $1000$ simulation runs (different random `seed') within each topology (total sample size is $5000$). The color bar represents the frequency of the NMSE data. Note that in some cases, our model fails to generate a meaningful NMSE as the reservoir output blows up. We get meaningful output from $\sim 90\% - 100\%$ cases depending on the reservoir sizes, and those data are plotted here and used to estimate the average NMSE. The bottom panel is the zoomed version of the top panel and the magenta dashed-dotted lines are the guide to the eye that shows the data distribution in the range of $\mu\pm\sigma$. The color codes to represent the $\mu$ and $\sigma$ are the same for the subsequent figures henceforth.}
    \label{fig2}
\end{figure*}

\begin{figure*}[htbp]
\begin{center}
\includegraphics[width=0.90\textwidth]{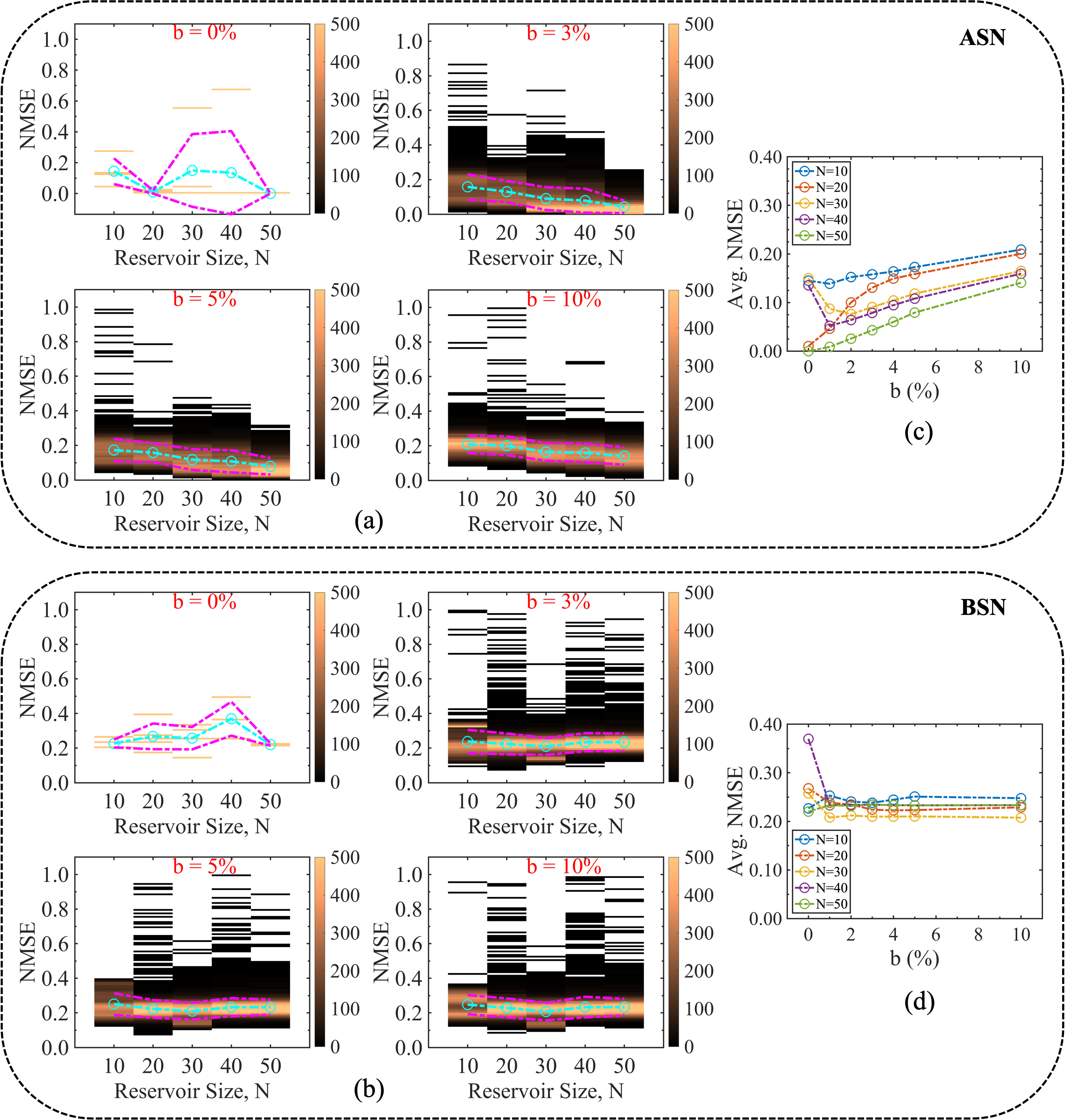}
\end{center}
    \caption{Evolution of NMSE for different degrees of stochasticity (noise percentages) associated with the (a) ASN and (b) BSN models. ASN performs better than the BSN model for analog time-series prediction tasks throughout the ranges of the degree of stochasticity as indicated by the average NMSE shown in (c) and (d) for ASN and BSN, respectively. The characteristics of the average NMSE as a function of reservoir size i.e., the decreasing trend for ASN while almost no change for BSN holds throughout the range of $b$.}
    \label{fig3}
\end{figure*}
\begin{figure*}[htbp]
\begin{center}
\includegraphics[width=0.88\textwidth]{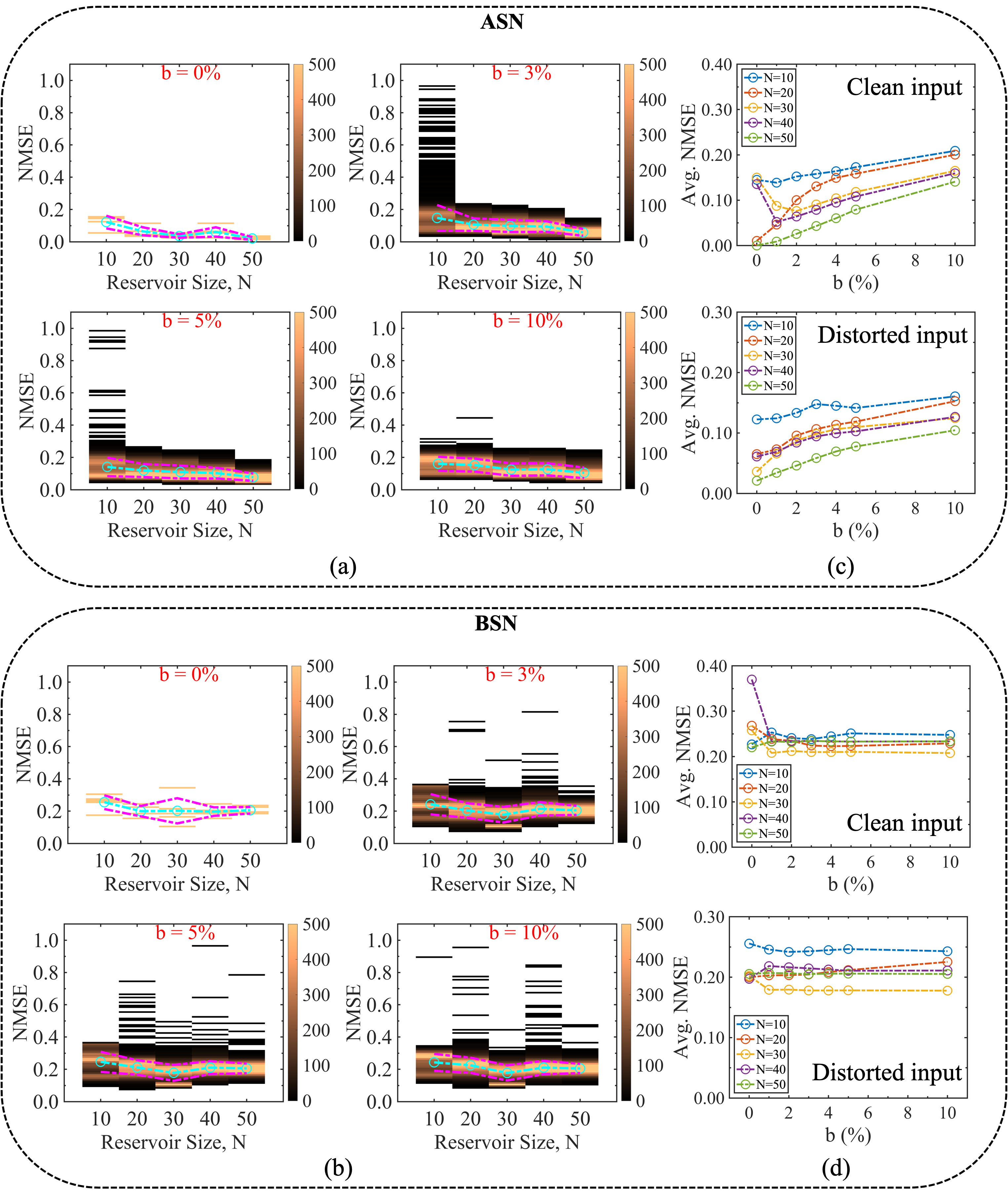}
\end{center}
    \caption{Evolution of NMSE for different degrees of stochasticity for (a) ASN and (b) BSN models for a distorted input signal. Random white noise is added to the clean input signal to introduce distortion and the form of the distorted signal is $u(t)=A \cos(2\pi f_1t)+B \sin(2\pi f_2t) + C[rand(1,t)-0.5]$, where $A=1$, $B=2$, $C=1$, $f_1=0.10~Hz$, and $f_2=0.02~Hz$. ASN performs better than BSN for the distorted input, as indicated by the average NMSE shown in (c) and (d) for ASN and BSN, respectively, which dictates the robustness of the ASN model in terms of performance irrespective of the input signals.}
    \label{fig4}
\end{figure*}
\section{Methods}
As discussed previously, the neuron models are analyzed in the context of a reservoir computer, specifically an echo-state network (ESN). An ESN is composed of a collection of recurrently connected neurons, with randomly distributed weights of the interconnects within this collection~\cite{ESN1,ESN2}. This forms the ``reservoir'', which is activated by an incoming signal, and whose output is read by an output layer trained via linear regression.

We employ different neuron models in this work, such as analog and binary neurons (with and without stochasticity in the model), which makes a total of four models at our disposal, namely, analog neuron (AN), analog stochastic neuron (ASN), binary neuron (BN), and binary stochastic neuron (BSN). The dynamical equations of the reservoirs built using different neuron models are described as follows~\cite{Samiran_ASN}:
\begin{eqnarray}
& \text{AN}: \textbf{x}[t+1]=(1-a)*\textbf{x}[t]+a*tanh(\textbf{z}[t+1]) \nonumber\\
& \text{ASN}: \textbf{x}[t+1]=(1-a)*\textbf{x}[t]+a*tanh(\textbf{z}[t+1])+\nonumber\\
& b*r_N[t]\nonumber\\
& \text{BN}: \textbf{x}[t+1]=(1-a)*\textbf{x}[t]+sgn(a*tanh(\textbf{z}[t+1]))\nonumber\\
& \text{BSN}: \textbf{x}[t+1]=(1-a)*\textbf{x}[t]+sgn(a*tanh(\textbf{z}[t+1])+\nonumber\\
& b*r_N[t])
\end{eqnarray}
where $\textbf{z}[t+1]=W^{in}\textbf{u}[t+1]+W^s\textbf{x}[t]$. Here, $\textbf{u}$ is the input vector, $\textbf{x}[t]$ represents the reservoir state vector at the time $t$, $a$ is the reservoir leaking rate (assumed to be the constant for all the neurons), $b$ is the neuron noise scaling parameter to include stochasticity in the neuron model, $r_N$ is a uniform random distribution, and $W^{in}$ and $W^s$ are the random weight matrices of input-reservoir and reservoir-reservoir connections, respectively. {We use the same leaking rate across all models to ensure a fair comparison among the neuron models on an equal footing. It can be challenging to compare models that have different parameters as it can introduce biases.} {One of the unique features of reservoir computing is having random weight matrices~\cite{RC1} and} we consider five different network topologies by creating five sets of $W^s$ using random `seed' for various reservoir sizes, {which makes our analysis unbiased to any particular network topology.} The $W^s$ elements are normalized using the spectral radius. We perform $1000$ simulations within each network topology making the total sample size $5000$ for every reservoir size within each neuron model. The output vector \textbf{y} is obtained as:

\begin{equation}
 \textbf{y} = W^{out}\textbf{x}
\end{equation}
where $W^{out}$ represents the reservoir-output weight matrix. We consider two different types of training methods, i.e, `offline' and `online' training. In the case of `offline' training, we extract the output weight matrix, $W^{out}$ once at the end of the training cycle and use that static $W^{out}$ for the testing cycle. In contrast, for `online' training, $W^{out}$ is periodically updated throughout the testing cycle. The entire testing cycle is divided into $40$ segments. The first segment uses the $W^{out}$ extracted from the initial training cycle. We calculate a new $W^{out}$ after the first segment of the testing cycle. Then, we update the $W^{out}$ such that the elements are composed of $90\%$ from the older version and $10\%$ from the new one. The updated $W^{out}$ is used for the second segment and the procedure keeps going on throughout the testing cycle. This stabilizes the learning at the cost of higher error rates as the learning evolution slowly evolves to a new configuration. This is akin to the successive over-relaxation methods used in many self-consistent numerical algorithms for improved convergence.
\begin{table*}[htbp]
\begin{center}
\caption{Average NMSE data extracted from the ASN and BSN models ($b = 5\%$ ) for various reservoir sizes. The form of the input signal is, $u(t)=A \cos(2\pi f_1t)+B \sin(2\pi f_2t) + C[rand(1,t)-0.5]$.\\}
\begingroup
\setlength{\tabcolsep}{10pt}
\renewcommand{\arraystretch}{1.5}
\begin{tabular}{|c|c|c|c|c|} 
\hline
Model & Reservoir Size & \multicolumn{3}{|c|}{Avg. NMSE for different input signals}\\
\hline
& & \begin{tabular}[c]{@{}c@{}}$\{A,~B,~C\}$ \\= $\{0.5,~1.0,~0.0\}$ \\ $\{f_1,~f_2\}$ \\= $\{0.20,~0.04\}~Hz$\end{tabular}  & \begin{tabular}[c]{@{}c@{}}$\{A,~B,~C\}$ \\= $\{1.0,~2.0,~0.5\}$ \\ $\{f_1,~f_2\}$ \\= $\{0.10,~0.02\}~Hz$\end{tabular}  & \begin{tabular}[c]{@{}c@{}}$\{A,~B,~C\}$ \\= $\{1.0,~2.0,~1.5\}$ \\ $\{f_1,~f_2\}$ \\= $\{0.10,~0.02\}~Hz$\end{tabular}\\
\hline
\multirow{5}{2em}{ASN} & $N=10$ & $0.1729$ & $0.1453$ & $0.1501$\\
& $N=20$ & $0.1585$ & $0.1199$ & $0.1161$\\
& $N=30$ & $0.1183$ & $0.0960$ & $0.0984$\\ 
& $N=40$ & $0.1080$ & $0.0775$ & $0.1001$\\
& $N=50$ & $0.0791$ & $0.0605$ & $0.0816$\\
\hline
\multirow{5}{2em}{BSN} & $N=10$ & $0.2510$ & $0.2396$ & $0.2546$\\
& $N=20$ & $0.2233$ & $0.2102$ & $0.2184$\\
& $N=30$ & $0.2103$ & $0.1895$ & $0.2028$\\ 
& $N=40$ & $0.2331$ & $0.2156$ & $0.2040$\\
& $N=50$ & $0.2329$ & $0.2142$ & $0.2173$\\ 
\hline
\end{tabular}
\endgroup
\label{table1}
\end{center}
\end{table*}
\begin{table*}[htbp]
\begin{center}
\caption{Robustness vs. accuracy trade-off ($N=20$). {The label Input $1$, Input $2$, Input $3$, and Input $4$ correspond to the sinusoidal clean input, sinusoidal with higher harmonic terms, sawtooth, and square input functions described earlier, respectively.}\\}
\begingroup
\setlength{\tabcolsep}{8.5pt}
\renewcommand{\arraystretch}{1.5}
\begin{tabular}{|c|c|c|c|c|c|c|c|c|c|} 
\hline
Model & b (\%) & \multicolumn{4}{|c|}{Blowup (\%)} & \multicolumn{4}{|c|}{Avg. NMSE}\\
\hline
& & \begin{tabular}[c]{@{}c@{}}Input 1\end{tabular}  & \begin{tabular}[c]{@{}c@{}}{Input 2}\end{tabular} & \begin{tabular}[c]{@{}c@{}}{Input 3}\end{tabular} & \begin{tabular}[c]{@{}c@{}}{Input 4}\end{tabular} & \begin{tabular}[c]{@{}c@{}}Input 1\end{tabular}  & \begin{tabular}[c]{@{}c@{}}{Input 2}\end{tabular} & \begin{tabular}[c]{@{}c@{}}{Input 3}\end{tabular} & \begin{tabular}[c]{@{}c@{}}{Input 4}\end{tabular}\\
\hline
AN & $0$ & $100$ & {$100$} & {$100$} & {$100$} & $-$ & {$-$} & {$-$} & {$-$}\\
\hline

\multirow{6}{2em}{ASN} & $1$ & $74.7$ & {$81.3$} & {$98.5$} & {$98.6$} & $0.3175$ & {$0.2759$} & {$0.4947$} & {$0.5475$}\\

&$2$ & $66.4$ & {$79.3$} & {$92.0$} & {$92.9$} & $0.2921$ & {$0.3225$} & {$0.3947$} & {$0.5537$}\\

&$3$ & $60.7$ & {$78.7$} & {$85.9$} & {$88.9$} & $0.2854$ & {$0.3301$} & {$0.3744$} & {$0.5591$}\\ 

& $4$ & $56.2$ & {$77.0$} & {$81.0$} & {$84.3$} & $0.2782$ & {$0.3534$} & {$0.3572$} & {$0.5515$}\\

& $5$ & $53.9$ & {$76.3$} & {$76.4$} & {$80.7$}& $0.2778$ & {$0.3597$} & {$0.3636$} & {$0.5358$}\\

& $10$ & $49.1$  & {$71.6$} & {$66.5$} & {$71.4$} & $0.2849$ & {$0.3903$} & {$0.3398$} & {$0.5316$}\\

& $15$ & $48.8$ & {$69.3$} & {$59.7$} & {$67.3$} & $0.3019$ & {$0.4266$} & {$0.3557$} & {$0.5412$}\\
\hline
\end{tabular}
\endgroup
\label{table2}
\end{center}
\end{table*}
\begin{table}[htbp]
\begin{center}
\caption{{Linear memory capacity (MC) for different neuron models.}\\}
{
\begingroup
\setlength{\tabcolsep}{10pt}
\renewcommand{\arraystretch}{1.5}
\begin{tabular}{|c|c|c|c|} 
\hline
Model & Reservoir Size & \multicolumn{2}{|c|}{MC}\\
\hline
& & \begin{tabular}[c]{@{}c@{}}$b = 0\%$\end{tabular}  & \begin{tabular}[c]{@{}c@{}}$b = 5\%$\end{tabular}\\
\hline
\multirow{2}{3em}{Analog} & $N=40$ & $39.0$ & $32.5$\\
& $N=50$ & $45.2$ & $36.2$\\
\hline
\multirow{2}{3em}{Binary} & $N=40$ & $2.7$ & $2.8$\\
& $N=50$ & $3.4$ & $3.2$\\
\hline
\end{tabular}
\endgroup
}
\label{table3}
\end{center}
\end{table}
\begin{figure*}[htbp]
\begin{center}
\includegraphics[width=.80\textwidth]{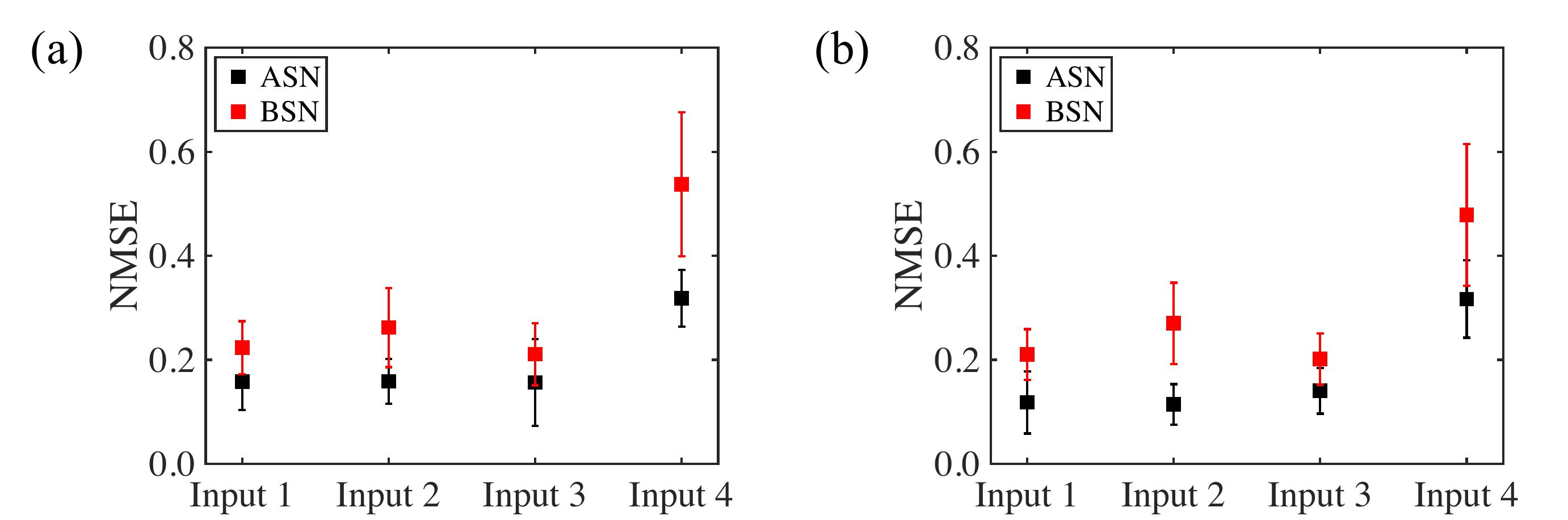}
\end{center}
    \caption{Comparison of NMSE for time-series prediction task between ASN and BSN models for various input functions for a reservoir size of (a) N = 20 and (b) N =30. The degree of stochasticity incorporated in both neuron models is $5\%$. The label Input $1$, Input $2$, Input $3$, and Input $4$ correspond to the sinusoidal clean input, sinusoidal with higher harmonic terms, sawtooth, and square input functions, respectively. ANS performance is better than BSN in terms of NMSE for different input functions.}
    \label{fig5}
\end{figure*} 

\begin{figure*}[htbp]
\begin{center}
\includegraphics[width=.75\textwidth]{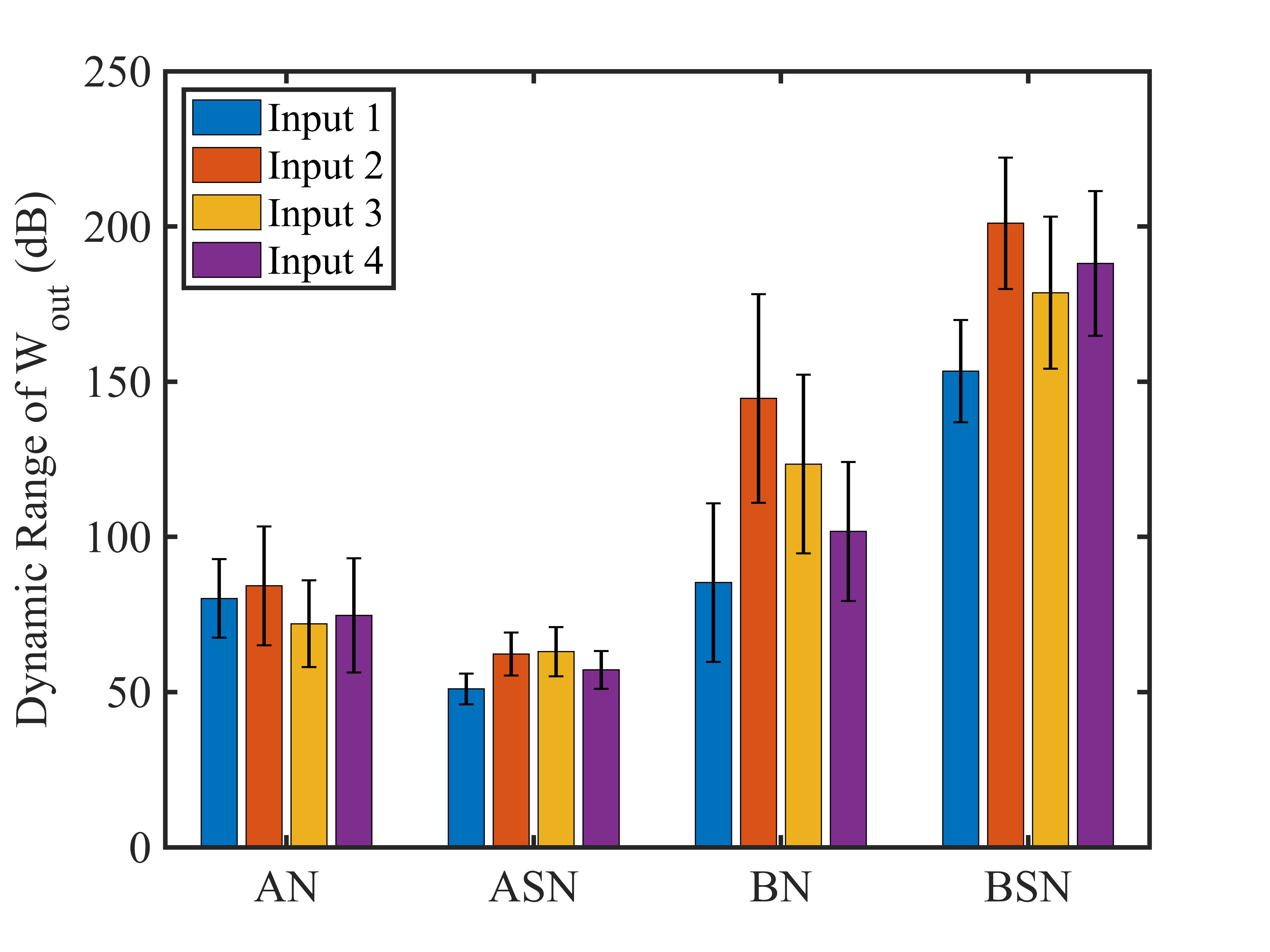}
\end{center}
    \caption{Dynamic range of the learned synaptic weights, $W_{out}$ for all the neuron models ($N = 20$). $5\%$ stochasticity is considered in the ASN and BSN models. ASN model shows the smallest dynamic range that leads to better hardware implementability. {The label Input $1$, Input $2$, Input $3$, and Input $4$ correspond to the sinusoidal clean input, sinusoidal with higher harmonic terms, sawtooth, and square input functions, respectively.}}
    \label{fig6}
\end{figure*}  
\section{Results and Discussions}
\subsection{Binary vs. Analog: Inference errors}
We implement the temporal inferencing task, specifically, the time-series prediction task to test and compare the performance of different neuron models. We consider an input signal of the form $u(t)=A \cos(2\pi f_1t)+B \sin(2\pi f_2t)$, which we referred to as a clean input. We use $A=1$, $B=2$, $f_1=0.10~Hz$, and $f_2=0.02~Hz$. {Although we choose the magnitude and frequency of the input arbitrarily, we further investigate other combinations of these variables (Table~\ref{table1}) to ensure that our analysis remains independent of them.} We train the neuron models using the clean input signal and test the models on a test signal from the same generator. The neuron models learn to reproduce the test signal from its previously self-generated output. The performance of the neuron models for time-series prediction tasks is usually measured by the NMSE, which is the metric that indicates how accurately the models can predict the test signal. If $y_{tar}$ is the target output and $y_{pre}$ is the actual predicted output, for $N_T$ time steps, we define NMSE as:
\begin{equation}
NMSE = \frac{1}{N_T(y_{tar}^{max}-y_{tar}^{min})}\sum_{i=1}^{i=N_T}{(y_{tar}(i)-y_{pre}(i))^2}
\end{equation}

Fig.~\ref{fig2}(a) and (b) show the NMSE for ASN and BSN, respectively for the time-series prediction task for various reservoir sizes. We generate the results using the `offline' training as discussed in the method section, for a clean input signal. We incorporate the stochasticity by adding $5\%$ white noise in both neuron models ($b=0.05$). The total sample size is $5000$ for a specific reservoir size, however, it is worth mentioning that we do not get valid NMSE for all the $5000$ cases because the network fails to predict the input signal and blows up for some cases. We get $\sim90\%-100\%$ successful cases depending on the reservoir sizes. Only valid data points are included in Fig.~\ref{fig2} and all the subsequent figures.  We find ASN performs better than BSN for all the reservoir sizes indicated by the average NMSE (cyan dashed-dotted line). Overall the NMSE is less scattered for ASN than BSN, so is their standard deviation, (magenta dashed-dotted line) as shown in the bottom panel of Fig.~\ref{fig2}. For ASN, we find that the average NMSE has a decreasing trend as the reservoir size increases, which indicates larger size networks can predict better. This happens because of the substantially richer dynamics and phase-space volume possible in a large network. In contrast, for BSN, the average NMSE is almost unchanged as the reservoir size increases.

We vary the stochasticity incorporated in the neuron models. Fig.~\ref{fig3}(a) and (b) show the distribution of the NMSE for different percentages of stochasticity, $b$ for ASN and BSN models, respectively. We find that ASN performs better than its BSN counterpart throughout the ranges of $b$ as indicated by the average NMSE. For ASN, the average NMSE shows a sub-linear trend as a function of $b$ (Fig.~\ref{fig3}(c)) for various reservoir sizes, while for BSN, the average NMSE remains unchanged (Fig.~\ref{fig3}(d)). For pure analog neuron ($b=0\%$), the NMSE is not much spread out, and also, for larger reservoir size, the average NMSE is smaller than the neuron model with stochasticity, however, having a neuron model with zero stochasticity is not practical. Moreover, stochasticity helps to make the system stable and reliable as discussed in the next section. Although the average NMSE increases with increasing $b$, we conjecture that $b=2-5\%$ would be optimal.

The aforementioned results are based on a clean input signal. We tested the models for distorted input as well. For the distorted case, we add a white noise in the clean input and the form of the distorted input signal is $u(t)=A \cos(2\pi f_1t)+B \sin(2\pi f_2t) + C[rand(1,t)-0.5]$. {The white noise is uniformly distributed for all $t$ values, both in the positive and negative half of the sinusoidal input. The degree of noise has been chosen arbitrarily. Again, we show various degrees of noise (Table~\ref{table1}) to make the analysis independent of a specific value of the noise margin.} The NMSE results shown in Fig.~\ref{fig4}(a) and (b) are calculated using $A=1$, $B=2$, $C=1$, $f_1=0.10~Hz$, and $f_2=0.02~Hz$. We find a better performance for ASN than that of BSN for the distorted input as well. It appears that for ASN, with a distorted input signal, the spectrum of NMSE is smaller, which reduces the standard deviation. The characteristics of the average NMSE are similar for the clean and distorted input for both ASN (Fig.~\ref{fig4}(c)) and BSN (Fig.~\ref{fig4}(d)) models. However, the average NMSE is slightly lower for the distorted input for both types of neuron models. Furthermore, we use different combinations of signal magnitude, frequency, and the weight of noise in the input signal. We list the average NMSE for various reservoir sizes in Table~\ref{table1}. {Additionally, we explore other input functions beyond the simple sinusoidal input used in the aforementioned results. In particular, we use a sinusoidal with higher harmonic terms, a sawtooth input function, and a square input function. The used form of the functions are $u(t)=\frac{4}{\pi}\sum_{i=1}^{15}{\frac{1}{n}\sin{2\pi nf_1t}}$ (odd $n$), $u(t)=A~sawtooth(2\pi f_1t)+B~sawtooth(2\pi f_2t)$, $u(t)=A~square(2\pi f_1t)+B~square(2\pi f_2t)$, respectively. In the case of sinusoidal with higher harmonic terms, we use the fundamental frequency $f_1 = 0.10~Hz$. For the sawtooth and square inputs, the magnitude and frequency remain the same as of the original sinusoidal clean input. The results are summarized in Fig.~\ref{fig5}, where the label Input $1$, Input $2$, Input $3$, and Input $4$ correspond to the sinusoidal clean input, sinusoidal with higher harmonic terms, sawtooth, and square input functions, respectively. Fig.~\ref{fig5} shows that for all the different inputs, ANS performance is better than BSN in terms of NMSE.} Comparing all the cases, we conjecture that ASN performs better than BSN for temporal inferencing tasks.

\subsection{Deterministic vs. Stochastic: Generalizability and Robustness}
One important aspect of any NN implementation is the generalizability and robustness of the learning. A model trained to a very specific data distribution will fail when it is running on a distribution that differs from the trained model. This is particularly true if a generative model guides its own subsequent learning, which is the example we have used in our online learning scenario. In this case, the underlying distribution is varied slowly while the network evolves its internal generative model to match the output of distribution, i.e., it works as a dynamically evolving temporal auto-encoder.

The stochasticity of the neuron response will add errors to the generated output as we see in the previous cases, however, we find that after a few iterations of the online learning cycle, the ability of this online learning blows up, i.e., the linear regression-based learning cannot keep up with the test distribution evolution and the error builds up (we call it blowup) and the whole training needs to be fully reset or reinitiated and cannot merely evolve from previous learning. This blowup occurs $100\%$ for deterministic analog neurons, and the rate reduces as the degree of stochasticity increases (parameter $b$). This is shown in Table~\ref{table2} {for various input functions}. It should be noted that at very high stochasticity while the training is more robust, the errors will be high, therefore a minimal amount of stochasticity is useful as a trade-off between these ends. The degree to which the trade-off can be performed depends on the application scenario. If full retraining is too expensive or not acceptable, then a relatively higher degree of stochasticity in the neuron is necessary, but if it is cheap and acceptable to retrain the whole network frequently, a near-deterministic neuron will be better suited to meet the requirements.

\subsection{Synaptic weights dynamic range: Hardware implementability}
One critical aspect of hardware implementability of neuromorphic computing is the ability to modulate the weights and the dynamic range or the order of magnitude to which weights may be distributed. It can be shown that a $30$-bit weight resolution represents about a $100~\mathrm{dB}$ dynamic range. While such ranges might be comparatively easily implemented in software, it is significantly difficult to implement such a high dynamic range in physical hardware. While some memristive materials may show multi-steps, it is hard to achieve much more than one order of magnitude change in the weights. Please note that we do not mean the change in the physical characteristics (typically the resistance) used to represent the weights themselves, but rather the number of steps that the weight can be implemented as.

We compare the dynamic range of the learned synaptic weights that need to be implemented in the reservoir networks (in the trained output readout layer) {for various input functions} and find that the ASN networks show the smallest dynamic range {for all the cases} (Fig.~\ref{fig6}) and suggest the easiest path to hardware implementability of physical neuromorphic computing. {It is important to note that the hardware implementation of neuromorphic computing is an open question and the dynamic range of the synaptic weights is one of the important factors when it comes to the physical deployment of neuromorphic computing as discussed above. ASN networks show better performance in terms of the dynamic range of learned synaptic weights compared to other models, which suggests that networks that employed ASN models might have better hardware implementability; however, it requires more analysis in terms of energy cost, scalability, and reconfigurability, which we leave as a future study.}

\subsection{{Memory Capacity}}

{The performance of reservoir computing is often described by memory capacity (MC)~\cite{MC1,MC2,MC3}. It measures how much information from previous input is present in the current output state of the reservoir. The task is to reproduce the delayed version of the input signal. For a certain time delay $k$, we measure how well the current state of the reservoir $y_k(t)$ can recall the input $u$ at time $t-k$. The linear MC is defined as:
\begin{equation}
MC = \sum_{k}{\frac{cov^2(u(t-k),y_k(t))}{\sigma^2(u(t-k)\sigma^2(y_k(t))}}
\end{equation}
where $u(t-k)$ is the delayed version of the input signal, which is the target output, and $y_k(t)$ is the output of the reservoir unit trained on the delay $k$. $cov$ and $\sigma^2$ denote $covariance$ and $variance$, respectively.}

{Table~\ref{table3} shows the linear MC for different neuron models for the distorted input $u(t)=A \cos(2\pi f_1t)+B \sin(2\pi f_2t) + C[rand(1,t)-0.5]$, where $A=1$, $B=2$, $C=1$, $f_1=0.10~Hz$, and $f_2=0.02~Hz$. We consider the delayed signal over $1$ to $50$ timesteps, meaning $k$ spans from $1$ to $50$. We find that Analog neurons have significantly larger linear MC than binary neurons. For analog neurons, linear MC increases as the reservoir size increases, which is expected because a larger dynamical system can retain more information from the past~\cite{MC1}. Additionally, including stochasticity in the analog neuron model degrades the linear MC as reported previously~\cite{MC1}. In contrast, binary neurons fail to produce substantial differences in linear MC when reservoir size is varied and stochasticity is included in the model.}

{Besides the previously mentioned properties, physical neuromorphic computing exhibits chaos or edge-of-chaos property, which has been shown to enhance the performance of complex learning tasks~\cite{chaos_1,chaos_2,chaos_3}. The edge-of-chaos property refers to the transition point between ordered and chaotic behavior in a system. In the discussed models, it may be possible to achieve the edge-of-chaos state by introducing increasing amounts of noise to the models, resulting in chaotic behavior that could potentially improve network performance. We find that with an increased degree of stochasticity in the neuron models, the learning process becomes more robust, which could be a signature of the performance improvement by including the edge-of-chaos property. However, the prediction accuracy and the linear MC tend to decrease with a higher degree of stochasticity, so the trade-off needs to be considered. It should be noted that a more comprehensive analysis is required to fully understand the impact of edge-of-chaos behavior on the discussed neuron models, which is beyond the scope of this paper and will be explored in future studies.}

\section{Conclusions}
In summary, we studied different neuron models for the analog signal inferencing (time-series prediction) task {in the context of reservoir computing} and evaluate their performances {for various input functions}. We show that the performance metrics are better for ASN than BSN for both clean and distorted input signals. We find that the increasing degree of stochasticity makes the models more robust, however, decreases the prediction accuracy. This introduces a trade-off between accuracy and robustness depending on the application requirements and specifications. Furthermore, the ASN model turns out to be the suitable one for hardware implementation, which attributes to the smallest dynamics range of the learned synaptic weights, {although other aspects, i.e., energy requirement, scalability, and reconfigurability need to be assessed}. {Additionally, we estimate the linear memory capacity for different neuron models, which suggests that analog neurons have a higher ability to reconstruct the past input signal from the present reservoir state.} These findings may provide critical insights for choosing suitable neuron models for real-time signal-processing tasks and pave the way toward building energy-efficient neuromorphic computing platforms.

\section*{Acknowledgments}
This work was supported by DRS Technology and in part by the NSF I/UCRC on Multi-functional Integrated System Technology (MIST) Center; IIP-1439644, IIP-1439680, IIP-1738752, IIP-1939009, IIP-1939050, and IIP-1939012. We thank Kerem Yunus Camsari, Marco Lopez, Tony Ragucci, and Faiyaz Elahi Mullick for useful discussions. All the calculations are done using the computational resources from High-Performance Computing systems at the University of Virginia (Rivanna) and the Extreme Science and Engineering Discovery Environment (XSEDE).
\bibliography{main}
\end{document}